\documentclass[12pt]{article}
\usepackage{amsmath}
\usepackage{amsfonts}
\usepackage[english]{babel}
\usepackage[ansinew]{inputenc}
\usepackage{graphicx}
\usepackage{graphpap}

\begin{document}

\newcommand{\bm}[1]{\mbox{\boldmath $#1$}}

\newtheorem{defi}{Definition}
\newtheorem{lema}{Lemma}
\newtheorem{proof}{Proof}
\newtheorem{thr}{Theorem}
\newtheorem{corollary}{Corollary}
\newtheorem{proposition}{Proposition}


\def\L{\mathcal{L}}
\def\Db{{\mathfrak D}_{b}}
\def\D{\mathfrak D}
\def\Sb{S_{b}}
\def\S{S}
\def\U{U}
\def\Sm{S_{2}}
\def\Sf{S_{f}}

\title{Stability of marginally outer trapped surfaces and symmetries}
\author{Alberto Carrasco$^1$ and Marc Mars$^2$ \\
Facultad de Ciencias, Universidad de Salamanca,\\
 Plaza de la Merced s/n, 37008 Salamanca, Spain. \\
$^1$ \, acf@usal.es, $^2$ \,marc@usal.es}

\maketitle

\begin{abstract}  We study properties of stable, strictly stable and
locally outermost marginally outer trapped surfaces in spacelike 
hypersurfaces of
spacetimes possessing certain symmetries such as isometries,
homotheties and conformal Killings.  We first obtain results for
general diffeomorphisms in terms of the so-called
metric deformation tensor and
then particularize to different types of symmetries. In
particular, we find restrictions at the surfaces on
the vector field generating the symmetry. Some consequences are discussed. As
an application we present a result on non-existence of stable
marginally outer trapped surfaces in slices of FLRW.
\end{abstract}

\section{Introduction}\label{sectionintroduction}

Trapped surfaces, and their various relatives, are fundamental objects in
Classical General Relativity. Being quasilocal versions of black holes,
their study is essential in order to 
understand how black holes evolve when no global assumptions are made
in the spacetime, for instance in order to address the cosmic
censorship conjecture (see e.g. \cite{Dafermos05}). They are also widely
used in numerical relativity. 

It is often the case that trapped surfaces (which will always be taken to be closed in this paper) 
have to be studied in 
spacetimes possessing some kind of symmetry. This is the case, for
instance, when configurations of equilibrium are considered, or in spherically 
symmetric or axially symmetric configurations. However, not only isometries are
important in this respect. For instance, critical collapse is a universal
feature of many matter models and the critical solution, which 
separates those configurations that disperse from those that form black holes,
is known to admit either a continuous
or a discrete self-similarity. This makes it interesting to study
trapped surfaces in spacetimes with homothetic Killing vectors. Many relevant spacetimes
admit other types of symmetries, like
for instance conformal symmetries, e.g. in FLRW cosmologies. Therefore, it 
becomes interesting to study the relationship between trapped surfaces and special types of 
vectors. A recent example of this interplay has been given in
\cite{BS09}, \cite{S08}, where the localization of the boundary of 
the set  containing trapped surfaces, which is a natural candidate for the "surface of an 
evolving black hole", was analyzed in the Vaidya spacetime, which is one of the simplest dynamical situations.
In this analysis the presence of a so-called Kerr-Schild 
symmetry (see e.g. \cite{CollHS01}) turned out to be fundamental. 

In the important case of isometries, general results on
the relationship between trapped surfaces and
Killing vectors were discussed in \cite{MS03}. The first variation of area was used to 
obtain several restrictions on the existence of trapped and marginally
trapped surfaces in spacetime regions possessing a causal Killing vector. More specifically,
if the Killing vector is timelike in some region, then no trapped surface can exist there,
and marginally trapped surfaces can only exist if their mean curvature vanishes identically. 
By obtaining a general identity for the first variation of area in terms of the 
deformation tensor of an arbitrary vector (see below for the definition)
similar restrictions were obtained for spacetimes admitting other types of symmetries, like
conformal Killing vectors or Kerr-Schild vectors. The same idea was also applied in \cite{S03b} 
to obtain analogous results in spacetimes with 
vanishing curvature invariants.

The interplay between isometries and 
dynamical horizons (which are spacelike hypersurfaces foliated
by marginally trapped surfaces) was considered in \cite{AG05} 
where it was proven that regular dynamical horizons cannot exist
in spacetime regions containing a nowhere vanishing causal Killing vector, provided the 
spacetime satisfies the null energy condition (NEC).

One of the most relevant variants of trapped surfaces 
are the so-called marginally {\it outer} trapped
surfaces (MOTS), where only the expansion along the outer null 
vector $\vec{l}$ becomes restricted. 
The relation between stable MOTS and isometries was 
considered in \cite{AMS08}, where
it was shown that, given a strictly stable MOTS $S$ 
in a hypersurface $\Sigma$ (not necessarily spacelike),
any Killing vector tangent
to $\Sigma$ on $S$ must in fact be tangent to $S$.

MOTS in stationary or static spacetimes play a particularly relevant role.
Indeed, MOTS are believed to be good replacements of black holes, so a 
natural question arises of whether or not some version of the
black hole uniqueness theorems also holds for asymptotically flat
equilibrium configurations containing MOTS. This was answered in the affirmative by P. Miao
\cite{Miao05}
in the static, vacuum case when the MOTS lies in a time symmetric slice (hence, it is a
minimal surface) and bounds a domain. A general study of MOTS in stationary and static
spacetimes with arbitrary matter contents satisfying NEC was performed in 
\cite{CM08}.  In the stationary case, it was proven that, on an arbitrary
spacelike hypersurface $\Sigma$, 
no bounding MOTS lying in the exterior region
where the Killing field $\vec{\xi}$ is causal can penetrate into the
timelike region. This result was strengthened for static Killing
vectors: no bounding MOTSs can penetrate in the exterior region where
the static Killing vector $\vec{\xi}$ is timelike. 
The underlying idea of \cite{CM08} was to take the outermost MOTS  $\S$ and construct another
weakly outer trapped surface which lies 
outside of $\S$, at least partially, 
thus contradicting the outermost property of $S$.
The new surface was constructed
by first moving $S$ to
the past of $\Sigma$ along the integral lines of the Killing vector $\vec{\xi}$ some 
amount $t$ and then
back to $\Sigma$ along the outgoing future null geodesics. The intersection of this 
hypersurface with $\Sigma$ defines a new surface $\S_t$, which
is automatically  
located partially outside of $S$ 
if the Killing is timelike somewhere on $S$. Furthermore,
the shift of $S$ along the isometry obviously gives a new MOTS, while 
the outer expansion cannot increase in the translation along the
null geodesics, due to the Raychaudhuri equation. Hence the whole procedure
gives a weakly outer trapped surface, and therefore a contradiction.

In the present work, we will study the interplay between 
stable and outermost properties of 
marginally outer trapped surfaces in spacetimes possessing special types of vector fields,
including isometries, homotheties, conformal Killing vectors and many others. In fact, we
will find several results involving completely general vector fields $\vec{\xi}$.
The initial idea is to analyze in detail the geometric construction of $S_t$ outlined
above in order to find restrictions on $\vec{\xi}$ on an outermost MOTS $S$ in a given
spacelike hypersurface $\Sigma$, or 
alternatively, forbid the existence of a MOTS in certain regions where 
$\vec{\xi}$ fails to satisfy those restrictions. 
The collection of $\{\S_t\}$ defines a variation of $S$ within $\Sigma$.
The corresponding first order variation of the outer null
expansion is an elliptic operator 
$L_m$ acting on a function $Q$  which 
is precisely the function which, at first order, determines whether $\S_t$ lies 
outside of $S$ or not. This observation, as such, is of little use until the operator can be directly
linked to the vector field $\vec{\xi}$, and more specifically, to its deformation tensor.
The standard expression for the stability operator (see e.g. \cite{AMS08}) has a priori
nothing to do with the properties of the vector field $\vec{\xi}$. The first task is, therefore, to
obtain an alternative (and completely general) expression for $L_m Q$ in terms of the deformation
tensor of $\vec{\xi}$. We devote Section \ref{variation} to do this. The result,
given in Proposition  \ref{propositionxitheta} below, is thoroughly used in this paper
and also has independent interest. 

With this expression at hand, we 
can already analyze under which conditions the procedure above gives restrictions on $\vec{\xi}$.
In Sect.\ref{LmQ} we concentrate on the case where $L_m Q$ has a sign everywhere on $S$. 
It turns out that the results obtained by the geometric construction above can, 
in most cases, be sharpened considerably by using the maximum principle of elliptic operators.
This also allows one to extend the validity of the results from the outermost case to 
the case of stable and strictly stable MOTS. The main result of Sect.\ref{LmQ} is given in Theorem
\ref{TrhAnyXi}, which holds for any vector field $\vec{\xi}$. This result is then particularized to 
conformal Killing vectors (including homotheties and Killing vectors). Under the additional 
restriction that the homothety or the Killing vector 
is everywhere causal and future (or past) directed, strong restrictions on the geometry of the MOTS
are derived (Corollary \ref{shear}). As a consequence, we prove that in a plane wave spacetime
any stable MOTS must be orthogonal 
to the direction of propagation of the wave. Marginally trapped surfaces are
also discussed in this section. 

As an explicit application of the results on conformal Killing vectors, 
we show, in subsection \ref{sectionFLRW}, that
MOTS which are stable with respect to any spacelike direction 
cannot exist in FLRW cosmological models provided
the density $\rho$ and pressure $p$ satisfy the inequalities $\rho \geq 0$, 
$\rho \geq 3 p $ and $\rho + p \geq 0$. This includes, for instance,
all classic models of matter and radiation dominant eras and also 
those models with accelerated expansion which satisfy NEC. Subsection \ref{sectiongeometric} deals
with one case where, in contrast with the standard situation, the geometric
construction does in fact give sharper results than the elliptic theory.

In the case when $L_m Q$ is not assumed to have a definite sign, the maximum principle
looses its power. However, the geometric construction can still be used despite the fact that 
the surfaces $S_t$ are necessarily {\it not} weakly outer trapped (for $t$ small 
enough). This is studied in Section \ref{sectionnonelliptic}, where we 
exploit a smoothing argument by 
Kriele and Hayward \cite{KH97} which allows one to construct, out of two intersecting surfaces,
a smooth surface which lies outside of them and 
has smaller outer expansion  than the original ones. This gives a result (Theorem
\ref{thrnonelliptic}) which holds for general vector fields $\vec{\xi}$ on any locally outermost MOTS.
As in the previous section, we then particularize to conformal Killing vectors, and then 
to causal Killing vectors and homotheties which, in this case, are
allowed to change their time orientation on $S$.

We start with the basic definitions and results needed for this work.

\section{Basics}
\label{sectionbasics}

Consider a spacetime $(M,g)$ and a
vector field $\vec{\xi}$ defined on it. The Lie derivative
$\L_{{\xi}}g_{\mu\nu}$ describes how
the metric is deformed along the local group of diffeomorphisms
generated by $\vec{\xi}$. We thus define the
{\bf metric deformation tensor} associated to
$\vec{\xi}$, or simply
deformation tensor, as
\begin{equation}\label{mdt}
a_{\mu\nu}\equiv
\nabla_{\mu}\xi_{\nu}+\nabla_{\nu}\xi_{\mu}.
\end{equation}
Special forms of $a_{\mu\nu}$ define special types of vectors. In particular,
$a_{\mu\nu}=2\phi g_{\mu\nu}$ ($\phi$ a scalar function)
defines a conformal Killing vector,
$a_{\mu\nu}=2 C g_{\mu\nu}$ ($C$ a constant) corresponds to
a homothety and $a_{\mu\nu}=0$ defines a Killing vector.

As described in the Introduction, we want to relate the deformation tensor
of special vectors to the stability and outermost properties of MOTS.
We will denote by $\S$ 
a smooth, closed (i.e. compact and without
boundary) and orientable surface embedded in a spacelike
hypersurface $\Sigma$. The future directed unit vector normal to $\Sigma$
will be called  $\vec{n}$
and the unit vector orthogonal to $\S$ along
$\Sigma$ is called $\vec{m}$. The null vectors
$\vec{l}=\vec{n}+\vec{m}$ and $\vec{k}=\vec{n}-\vec{m}$ are
a null basis of the normal bundle of $\S$, and satisfy
$( \vec{l} \cdot \vec{k})=-2$ (scalar product with the spacetime metric
is denoted by $( \, \cdot \, )$). These vectors are univocally defined
once a choice of orientation for $\vec{m}$ is made.

The first fundamental form of $S$ is a Riemannian metric which we denote by
$\gamma_{AB}$. At any point $p \in S$, the tangent space $T_p M$
decomposes as the direct sum of the tangent and normal vector spaces to $\S$.
This splits any vector $\vec{V} \in T_p M$ as
$\vec{V} = \vec{V}^{\parallel} + \vec{V}^{\perp}$.
The second fundamental form vector of $\S$
is defined as
$\vec{\kappa}_{AB}\equiv -{\left(
\nabla_{\vec{e}_{A}}\vec{e}_{B}\right)^{\perp}}$ where
$\{\vec{e}_{A} |_p \}$ is a basis of $T_p \S$. Finally, the
mean curvature vector is the trace of the second fundamental form,
$\vec{H}=\gamma^{AB}\vec{\kappa}_{AB}$. Being a normal vector, it can
be expanded in the null basis as
\[ \vec{H}=-\frac12\left( \theta_{l}\vec{k} +
\theta_{k}\vec{l} \, \right),
\]
where the coefficients define the
null expansions $\theta_{l}$, $\theta_{k}$ of
$\S$ along $\vec{l}$ and $\vec{k}$, respectively. Similarly, 
the expansion along any normal direction $\vec{\eta}$ is defined
as $\theta_{\eta} \equiv ( \vec{H} \cdot \vec{\eta} )$
and the second fundamental form along $\vec{\eta}$ is
$\kappa^{\eta}_{AB} = ( \vec{\kappa}_{AB} \cdot \vec{\eta} )$.

A useful classification of surfaces arises depending on the causal
character of $\vec{H}$ or on the sign of one of the expansions.
Assume that one preferred orientation of $\vec{m}$
can be selected geometrically. We 
call this the {\it outer} direction. The corresponding null vector
$\vec{l} = \vec{n} + \vec{m}$ is the outer null direction.
If, furthermore,
$\S$ separates $\Sigma$ in two regions, we will call ``exterior'' the portion to which
the outer direction points.

The types of surfaces
that will play a role in this paper are (see \cite{S07}
for an exhaustive classification):
$\S$ is {\bf marginally future (past) trapped} if $\vec{H}$ points
along one of the null normals, $\vec{l}$ or $\vec{k}$, and is future (past) pointing at each point
(in our convention, the vanishing vector is both future and past null),
$\S$ is {\bf weakly outer trapped} if $\theta_{l} \leq 0$ and
$S$ is {\bf marginally outer trapped surface (MOTS)} provided $\theta_{l}=0$.

This work basically deals with properties of
(strictly) stable and locally outermost MOTSs, defined as follows
\cite{AMS05}.  $\S$ is {\bf stable}\footnote{Strictly speaking we should
say stable {\it in $\Sigma$}. However, we will only deal with one hypersurface
at a time, and no confusion should arise.}
if there exists a function $\psi\geq 0$,
$\psi\not\equiv 0$ on $\S$ such that the variation of
$\theta_{l}$ along $\psi\vec{m}$, denoted by
$\delta_{\psi{m}}\theta_{l}$, is non-negative.  $\S$ is
{\bf strictly stable} if, moreover, $\delta_{\psi
m}\theta_{l}\neq 0$ somewhere on $\S$. As described in
\cite{AMS05}, the variation
$\delta_{\psi{m}}\theta_{l}$ gives a linear
second order elliptic operator acting on $\psi$, which we will denote by
$L_{m}\psi$. The explicit form of this operator appears in
equation (1) of \cite{AMS05}. It is also well-known \cite{AMS05, AMS08}
that stability can be rephrased in
terms of the sign of principal eigenvalue
$\lambda_{m}$  of $L_m$ (defined to have the smallest real part, and which 
is always real): $S$ is stable if
$\lambda_{m} \geq 0$ and strictly stable if
$\lambda_{m} >0$.

A MOTS  $\S$ is {\bf locally outermost} if there exists a
two-sided neighbourhood of $\S$ on $\Sigma$ whose exterior part does not contain
any weakly outer trapped surface. We will denote by $\mathfrak{D}$ the interior part of this two-sided neighbourhood.

The relationship between these types of surfaces is the following \cite{AMS05}:
(i) a strictly stable MOTS is necessarily locally outermost, (ii) a locally outermost
MOTS is necessarily stable, and (iii) none of the converses is true
in general.

The results obtained in Sect.\ref{LmQ} below use the following version of the
maximum principle for second order
linear elliptic operators \cite{AMS08} (recall that the eigenspace
corresponding to the principal eigenvalue is one-dimensional and no function in this space
can change sign).
\begin{lema}\label{lemmaelliptic}
Consider a second order linear elliptic operator $L$ on a
compact manifold $\S$ with principal eigenvalue $\lambda\geq 0$
and principal eigenfunction $\phi$ and let $\psi$ be a smooth function 
satisfying $L \psi \geq 0$ ($L \psi \leq 0$).
\begin{enumerate}
\item If $\lambda=0$, then $L \psi \equiv 0$ and $\psi=C\phi$ for some constant
$C$

\item If $\lambda>0$ and $L \psi \not\equiv 0$, then $\psi>0$ ($\psi<0$) all over
$\S$.
\item If $\lambda>0$ and $L \psi \equiv 0$, then $\psi\equiv 0$.
\end{enumerate}
\end{lema}

As mentioned in the Introduction, the idea we want to apply in order
to obtain restrictions on a given vector field $\vec{\xi}$ on a MOTS $\S$
consists in moving $\S$ first along the integral lines of
$\vec{\xi}$ a parametric amount $t$. This gives a new surface $S^{\prime}_t$.
Take the
null normal $\vec{l}^{\prime}_t$ on this surface
which coincides with the continuous deformation of
$\vec{l}$ and consider the null hypersurface generated by null geodesics
with tangent vector $\vec{l}^{\prime}_t$. This hypersurface is smooth close
enough to $\S^{\prime}_t$. Being null, its intersection with the spacelike
hypersurface $\Sigma$ is transversal and hence defines a smooth surface
$S_t$ (for $t$ sufficiently small). By this construction, a point $p$ on $S$
describes a curve in $\Sigma$. The tangent vector of
this curve on $\S$, denoted by $\vec{\nu}$,
will define the variation vector generating the
deformation $\{ \S_t \}$ of $S$. Figure 1 gives a graphic representation of this 
construction.

\begin{figure}
\begin{center}
\includegraphics[width=9cm]{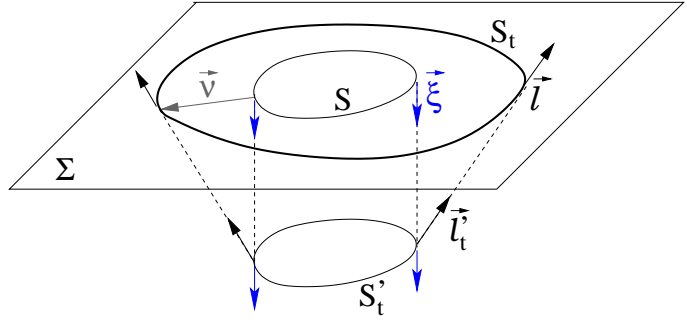}
\caption {The figure represents how the new surface $S_t$ is constructed
from the original surface $S$. The intermediate surface $S^{\prime}_t$ is
obtained from $S$ by
dragging along $\vec{\xi}$ a parametric amount $t$.
}
\end{center}
\end{figure}

As usual, we decompose the vector $\vec{\xi}$ into normal and tangential
components with respect to $\Sigma$, as
$\vec{\xi}=N\vec{n}+\vec{Y}$. On $S$ we will further decompose
$\vec{Y}$ in terms of a tangential component $\vec{Y}^{\parallel}$,
and a normal component $(\vec{Y} \cdot \vec{m}) \vec{m}$, i.e.
$\vec{\xi} |_S =N_S \vec{n}+ (\vec{Y} \cdot \vec{m} ) \vec{m}  +
\vec{Y}^{\parallel}$, where $N_S$ is the value of $N$ on the surface.
Since $\vec{\nu}$ defines the variation of $\S$
to first order, we only need to evaluate the vector $\vec{l}^{\prime}_t$ to
zero order in $t$, which obviously coincides with $\vec{l}$. It follows
that $\vec{\nu}$ is a linear combination (with functions) of $\vec{\xi} |_{S}$
and $\vec{l}$. The amount
we need to move $S^{\prime}_t$ in order to go back to $\Sigma$ can
be determined by imposing $\vec{\nu}$ to be tangent to $\Sigma$. This
gives $\vec{\nu} =  \vec{\xi} - N_S \vec{l} = Q\vec{m}+\vec{Y}^{\parallel}$,
where
\begin{equation}\label{Q}
Q= ( \vec{Y} \cdot \vec{m} ) -N_S = ( \vec{\xi} \cdot \vec{l} \, ),
\end{equation}
Since the tangential part of $\vec{\nu}$ does not
affect the variation of $\theta_{l}$ along $\vec{\nu}$ for a MOTS,
it follows that $\delta_{{\nu}}\theta_{l}=L_{m}Q$. Then, a
direct application of
Lemma \ref{lemmaelliptic} for a MOTS $\S$ with stability operator
$L_m$ leads to the following result.
\begin{lema}
\label{lemmaelliptic2}
Let $\S$ be a stable MOTS on a spacelike hypersurface $\Sigma$.
If $\left. L_{m}Q\right|_{\S}\leq 0$ ($\left.L_{m}Q\right|_{\S}\geq 0$) and not identically zero, 
then $\left. Q\right|_{\S}<0$ ($\left. Q\right|_{\S}>0$).

Furthermore, if $\S$ is strictly stable and 
$\left. L_{m}Q\right|_{\S}\leq 0$ ($\left.L_{m}Q\right|_{\S}\geq 0$) then 
$\left. Q\right|_{\S}\leq 0$ ($\left. Q\right|_{\S}\geq 0$) and it vanishes at one 
point only if it vanishes everywhere on $\S$.
\end{lema}
This result will be used in Sect.\ref{LmQ} to obtain restrictions
on the vector field $\vec{\xi}$ on stable and strictly stable MOTS.
The idea is to use the deformation tensor
to obtain an independent expression for $L_m Q$. Consider the simplest
example of a Killing vector $\vec{\xi}$. Since the null expansion does not
change under an isometry, it follows that the surface $S^{\prime}_t$ is
also a MOTS. Moving back to $\Sigma$ along the null hypersurface
gives a contribution to  $\theta_l (S_t)$ which, from
the Raychaudhuri equation, is easily computed
to be $L_m Q = N_S W$, where
we have introduced the shorthand notation
\begin{equation}\label{W}
W=\kappa_{l}^{2}+G_{\mu\nu}l^{\mu}l^{\nu},
\end{equation}
with $G_{\mu\nu}$ being the Einstein tensor of $(M,g)$ and
$\kappa_l^2=\kappa^{l}_{AB}
\kappa^{l \, AB}$ the square of the second fundamental form along $\vec{l}$, which 
coincides with the square of the shear along $\vec{l}$ in the case of MOTS.
Note that $W$ is non-negative
provided the null energy condition
(NEC) holds. i.e. $G_{\mu\nu}w^{\mu}w^{\nu} \geq 0$  for any null vector
$\vec{w}$. It is clear that under NEC Lemma \ref{lemmaelliptic2}
implies restrictions on any Killing vector on a stable MOTS.

However, obtaining the result $L_m Q = N_S W$ directly
from the explicit form of
the elliptic operator $L_m$ 
is not trivial because
the condition  of $\vec{\xi}$ being a Killing vector does not give
obvious restrictions on the coefficients of this operator. In the case of
Killing vectors, the point of view of moving along $\vec{\xi}$ and then back to
$\Sigma$ gives a simple method of calculating
$L_m Q$. For more general vectors, however, the motion along $\vec{\xi}$
will give a non-zero contribution to $\theta_{l}$ which needs
to be computed (for Killing vectors this term was known to be zero
via a symmetry argument, not from
a direct computation). In order to do this, it becomes necessary
to have an alternative, and completely general, expression for
$\delta_{\xi} \theta_l$ directly in terms of the deformation tensor $a_{\mu\nu}$
of $\vec{\xi}$.

\section{Variation of the expansion and the metric deformation tensor}
\label{variation}

The aim of this section is to derive an identity for $\delta_{\xi} \theta_l$
in terms of $a_{\mu\nu}$. This result will be important later on in this paper,
and may also be of
independent interest. We derive this expression
in full generality, i.e. without assuming $\S$ to be a MOTS and for the expansion
$\theta_{\eta}$ along any normal vector $\vec{\eta}$ of $\S$, not necessarily a null normal.

To do this calculation, we need to take derivatives of tensorial objects defined
on each one of $\S^{\prime}_t$. For a given point $p \in \S$, these tensors live on different spaces,
namely the tangent spaces of $\phi_t (p)$, where $\phi_t$ is the local diffeomorphism generated by $\vec{\xi}$.
In order to define the variation, we need to pull-back all tensors to the point $p$ before doing the derivative.
We will denote the resulting derivative by $\L_{{\xi}}$. This is, in fact, an abuse of notation because
we are not taking Lie derivatives of tensor fields on the manifold (they are tensorial objects on each $S^{\prime}_t$
but these surfaces may perfectly well intersect each other).
Nevertheless,
it is a useful notation because when acting on spacetime tensor fields
(e.g. the metric $g$) the operation
involved is really the standard Lie derivative along $\vec{\xi}$. This will simplify the calculation considerably.

Notice in particular that the definition of $\theta_{\eta}$ depends on the choice of $\vec{\eta}$
on each of the surfaces $S^{\prime}_t$. Thus $\delta_{\xi} \theta_{\eta}$ will necessarily 
include a term of the form $\L_{\xi} \eta_{\alpha}$ which is not uniquely defined
(unless $\vec{\eta}$ can be uniquely defined on each $S^{\prime}_t$ which is usually not
the case). Nevertheless, for the
case of MOTS and when $\vec{\eta} = \vec{l}$ this a priori ambiguous term
becomes determined, as we will see. The general expression
for $\delta_{{\xi}} \theta_{\eta}$ is given in the following proposition.

\begin{proposition}\label{propositionxitheta} Let $\S$ be a surface on a spacetime $(M,g)$,
$\vec{\xi}$ a vector field defined on $M$ with deformation tensor $a_{\mu\nu}$ and
$\vec{\eta}$ a vector field normal to $\S$.  Then, the variation along $\vec{\xi}$
of the expansion $\theta_{\eta}$ on $\S$ reads
\begin{eqnarray}\label{xithetau}
\delta_{{\xi}}\theta_{\eta}&=&
H^{\mu}\L_{{\xi}}\eta_{\mu}-a_{AB}
\kappa^{AB}_{\mu}\eta^{\mu}
\left .
+ \gamma^{AB}e_{A}^{\alpha}e_{B}^{\rho}\eta^{\nu}
	\left[ \frac{1}{2}\nabla_{\nu}a_{\alpha\rho} - \nabla_{\alpha}a_{\nu\rho}\right]
\right|_{\S}.
\end{eqnarray}
where $a_{AB}\equiv e_{A}^{\alpha}e_{B}^{\beta}a_{\alpha\beta}$.
\end{proposition}

{\bf Proof.} Since $\theta_{\eta}=H^{\mu}\eta_{\mu}=\gamma^{AB}\kappa_{AB}^{\mu}\eta_{\mu}$, the
variation we need to calculate involves three terms
\begin{equation}\label{Lietheta}
\delta_{\xi}\theta_{\eta}=\L_{\xi}\gamma^{AB}
\kappa_{AB}^{\mu}\eta_{\mu} + \gamma^{AB}\L_{\xi}\kappa_{AB}^{\mu}\eta_{\mu} + H^{\mu}\L_{\xi}\eta_{\mu}.
\end{equation}
In order to do the calculation, we will choose $\phi_t^{\star} (\vec{e}_A)$ as
the basis of tangent vectors at $\phi_t(p) \in S^{\prime}_t$. This entails no loss
of generality and implies  $\L_{\xi}\vec{e}_{A}=0$, which makes the calculation simpler.
Our aim is to express each term of (\ref{Lietheta}) in terms of $a_{\mu\nu}$. For the first term,
we need to calculate  $\L_{\xi}\gamma^{AB}$. We start with
$\L_{\xi}\gamma_{AB}=\L_{\xi} \left ( g (\vec{e}_A, \vec{e}_B ) \right ) =
( \L_{\xi} g ) \left (\vec{e}_A,\vec{e}_B \right ) =
a_{\mu\nu} e_{A}^{\mu}e_{B}^{\nu}\equiv
a_{AB}$, which immediately implies
$\L_{\xi}\gamma^{AB}=-a_{CD}\gamma^{AC}\gamma^{BD}$,
so that the first term in (\ref{Lietheta}) becomes
\begin{equation}\label{firstterm}
\L_{\xi}\gamma^{AB} \kappa_{AB}^{\mu}\eta_{\mu}=
- a_{AB}\kappa^{AB}_ {\mu}\eta^{\mu},
\end{equation}
where capital Latin indices are lowered and raised with $\gamma_{AB}$ and
its inverse.

The second term $\gamma^{AB}(\L_{\xi}\kappa_{AB}^{\mu}) \eta_{\mu}$ is more complicated.
It is useful to introduce the projector to the normal space of $S$,
$h_{\nu}^{\mu}\equiv \delta^{\mu}_{\nu} - g_{\nu\beta}e_{A}^{\mu}e_{B}^{\beta}\gamma^{AB}$.
From the previous considerations, it follows that
$\L_{\xi} h_{\nu}^{\mu}=e_{A}^{\mu}e_{B}^{\beta}
	(  a^{AB} g_{\nu\beta} - \gamma^{AB}a_{\nu\beta})$, which implies 
\begin{eqnarray}
\L_{\xi} (\kappa^{\mu}_{AB} ) \eta_{\mu} = -
\L_{\xi} \left ( h^{\mu}_{\nu} e_{A}^{\alpha}\nabla_{\alpha}e_{B}^{\nu} \right)
\eta_{\mu} = -
\eta_{\nu}\L_{\xi}\left( e_{A}^{\alpha}\nabla_{\alpha}e_{B}^{\nu}\right).
\label{LiePi}
\end{eqnarray}
where we have used the fact that $\eta_{\mu}$ is orthogonal to $\S$, so its contraction
with $\L_{\xi} h^{\mu}_{\nu}$ vanishes.

Therefore we only need to evaluate
$\L_{\xi} \left( e_{A}^{\alpha}\nabla_{\alpha}e_{B}^{\nu}\right)$. It is well-known
(and in any case easily verifiable) that for an arbitrary
vector field $\vec{v}$, the commutation
of the covariant derivative and the Lie derivative introduces a term involving the Riemann
tensor $R^{\nu}_{\,\,\rho\sigma\alpha}$ of $g$, as follows
\[
\L_{\xi}\nabla_{\alpha}v^{\nu}-\nabla_{\alpha}\L_{\xi}v^{\nu}
=v^{\rho}\nabla_{\alpha}\nabla_{\rho}\xi^{\nu}
+R^{\nu}_{\,\,\,\rho\sigma\alpha}v^{\rho}\xi^{\sigma},
\]
This expression is still true for the variational derivative we are calculating.
Thus, we have
\begin{eqnarray}
\L_{\xi}\nabla_{\alpha}e_{B}^{\nu}
=e_{B}^{\rho}\nabla_{\alpha}\nabla_{\rho}\xi^{\nu}
+R^{\nu}_{\,\,\,\rho\sigma\alpha}e_{B}^{\rho}\xi^{\sigma}. \label{Liee}
\end{eqnarray}
It only remains to express the quantity $\nabla_{\alpha}\nabla_{\rho}\xi^{\nu}
+R^{\nu}_{\rho\sigma\alpha}\xi^{\sigma}$ in terms of $a_{\mu\nu}$. To that aim,
we take a derivative of equation (\ref{mdt}) and use the Ricci
identity to get
\[
\nabla_{\nu}\nabla_{\alpha}\xi_{\rho} + \nabla_{\alpha}\nabla_{\rho}\xi_{\nu}=
R_{\sigma\rho\alpha\nu}\xi^{\sigma} + \nabla_{\alpha}a_{\nu\rho}.
\]
Now, write the three equations obtained from this one by cyclic
permutation of the three indices.
Adding two of them and subtracting the third one we find, after using the
first Bianchi identity,
\[
\nabla_{\alpha}\nabla_{\rho}\xi_{\nu} =
R_{\sigma\alpha\rho\nu}\xi^{\sigma}
+
\frac12\left[ \nabla_{\alpha}a_{\nu\rho}+\nabla_{\rho}a_{\alpha\nu}-\nabla_{\nu} a_{\alpha\rho} \right].
\]
Substituting (\ref{Liee}) and this expression into (\ref{LiePi}) yields
\begin{equation}\label{secondterm}
\gamma^{AB}\L_{\xi}\kappa_{AB}^{\mu}\eta_{\mu}=
	 \gamma^{AB}e_{A}^{\alpha}e_{B}^{\rho}\eta^{\nu}
	\left[ \frac{1}{2}\nabla_{\nu}a_{\alpha\rho} - \nabla_{\alpha}a_{\nu\rho}\right].
\end{equation}
Inserting (\ref{firstterm}) and (\ref{secondterm})
into equation (\ref{Lietheta})  proves the lemma.
$\hfill \square$.

We can now particularize 
to the outer null expansion in a MOTS.
\begin{corollary}
\label{corollaryxitheta}
If $S$ is a MOTS then
\begin{eqnarray}\label{xitheta} \delta_{{\xi}}\theta_{l}
&=&-\frac14\theta_{k}a_{\mu\nu}l^{\mu}l^{\nu}-a_{\mu\nu}e_{A}^{\mu}
e_{B}^{\nu}\kappa^{AB}_{\rho}l^{\rho}
\left .
+ \gamma^{AB}e_{A}^{\alpha}e_{B}^{\rho} l^{\nu}
	\left[ \frac{1}{2}\nabla_{\nu}a_{\alpha\rho} - \nabla_{\alpha}a_{\nu\rho}\right]
\right|_{\S}.
\end{eqnarray}
\end{corollary}

{\bf Proof.}  The normal vector $\vec{l}_t^{\prime}$ defined on each of the surfaces $S^{\prime}_t$ is null.
Therefore, using $\mathcal{L}_{\xi} \, g^{\mu\nu} = - a^{\mu\nu}$,
\begin{eqnarray}
0 = \mathcal{L}_{\xi} \left ( {l_t}^{\prime}_{\mu} {l_t}^{\prime}_{\nu} g^{\mu\nu} \right ) = 2 l^{\mu} \mathcal{L}_{\xi}  
{l_t}^{\prime}_{\mu}
- a_{\mu\nu} l^{\mu} l^{\nu}.
\label{null}
\end{eqnarray}
Since, on a MOTS  $\vec{H}=-\frac12 \theta_{k}\vec{l}$, it follows
$H^{\mu} \mathcal{L}_{\xi} {l_t}^{\prime}_{\mu} = -\frac{1}{2} \theta_k l^{\mu} \mathcal{L}_{\xi} \, {l_t}^{\prime}_{\mu} =
- \frac{1}{4} \theta_k a_{\mu\nu} l^{\mu} l^{\nu}$, and the corollary follows
from (\ref{xithetau}).  $\hfill \square$

{\bf Remark.} From the proof, it is clear that we have only used $\theta_l =0$ at $p$. Therefore
formula (\ref{xitheta}) holds in general for arbitrary surfaces $\S$ 
at any point where $\theta_l=0$.

\section{Results provided $L_m Q$ has a sign on $\S$}
\label{LmQ}

The most favorable case to obtain restrictions on the generator $\vec{\xi}$
on a given MOTS $S$ is when the surfaces $\{ S_t\} $ constructed
by the procedure above are weakly outer trapped.
This is guaranteed for small enough $t$ when $L_m Q$ is strictly
negative everywhere, because then this first order term becomes
dominant. Suppose that in addition of being a MOTS $S$ is also {\it outermost},
in the intuitive sense
that no other weakly outer trapped surface can penetrate in its exterior (we will give a more
precise definition below).  Since the direction to which a point $p \in S$
moves  is determined to first order by the vector $\vec{\nu} = Q \vec{m} +
\vec{Y}^{\parallel}$, it is clear that $Q >0$ at any point implies that for small
enough $t$, $S_t$ lies partially in the exterior of $S$. Combining these facts, it follows
that $L_mQ <0$ everywhere and $Q>0$ somewhere is impossible for an outermost MOTS.
This argument is intuitively very clear. However, this geometric method does not provide the most powerful
way of finding this type of restrictions. Indeed, when the first order term $L_m Q$
vanishes at some points, then higher order coefficients come necessarily into play, which makes the geometric
argument involved.  It is remarkable that using the elliptic results described in Sect. \ref{sectionbasics},
most of these situations can be treated in a satisfactory way. Furthermore, since the elliptic methods
only use infinitesimal information, there is no need to restrict oneself to outermost MOTS,
and the more  general case of stable or strictly stable surfaces can be considered.
In this section we will give several results along these lines. The general idea is to
combine Lemma \ref{lemmaelliptic2} with the general calculation for the variation
of $\theta_l$ obtained in the previous section to get restrictions on special types of generators $\vec{\xi}$
on a stable or strictly stable MOTS.

Our first result is fully general in the sense that it is valid for any generator $\vec{\xi}$.
\begin{thr}
\label{TrhAnyXi}
Let $S$ be a stable MOTS on a spacelike hypersurface $\Sigma$ and $\vec{\xi}$ a vector field
on $S$ with deformation tensor $a_{\mu\nu}$. With the notation above, define
\begin{eqnarray}
Z = -\frac14\theta_{k}a_{\mu\nu}l^{\mu}l^{\nu}-a_{AB}\kappa^{AB}_{\mu}l^{\mu}
\left .
+ \gamma^{AB}e_{A}^{\alpha}e_{B}^{\rho} l^{\nu}
	\left[ \frac{1}{2}\nabla_{\nu}a_{\alpha\rho} - \nabla_{\alpha}a_{\nu\rho}\right]
+ N W \right|_{\S},
\label{Z}
\end{eqnarray}
and assume $Z \leq 0$ everywhere on $S$.
\begin{itemize}
\item[(i)] If $Z \neq 0$ somewhere, then $(\vec{\xi} \cdot \vec{l} ) < 0$ everywhere.
\item[(ii)] If $S$ is strictly stable, then $(\vec{\xi} \cdot \vec{l} ) \leq 0$
everywhere and vanishes at one point only if it vanishes everywhere.
\end{itemize}
\end{thr}

{\bf Remark.} The theorem also holds if all the inequalities are reversed. This follows
directly by replacing $\vec{\xi} \rightarrow - \vec{\xi}$.

{\bf Proof.}
Consider the
first variation of $S$ defined by the vector $\vec{\nu} = \vec{\xi} - N_S \vec{l} =
Q \vec{m} + \vec{Y}^{\parallel}$. From the definition of stability operator \cite{AMS08}, we have
$\delta_{\nu} \theta_l = L_m Q$. On the other hand, linearity of this variation  gives
$\delta_{\nu} \theta_l = \delta_{\xi} \theta_{l} - N_S \delta_{l} \theta_{l}$.
Using now the Raychaudhuri equation $\delta_{l} \theta_{l} = - W$ (see (\ref{W})) and the
identity (\ref{xitheta}) gives $L_m Q = Z$. Since $Q = ( \vec{\xi} \cdot \vec{l} )$,
the result follows directly from Lemma \ref{lemmaelliptic2}. $\hfill \square$

This theorem gives information about the relative position between the generator
$\vec{\xi}$ and the outer null normal $\vec{l}$ and has, in principle, many potential consequences.
Specific applications require considering spacetimes
having special vector fields for which sufficient information about its deformation tensor is available.
Once such a vector is known to exist, the result above can be used either to restrict the
form of $\vec{\xi}$ in stable or strictly stable MOTS or, alternatively, to
restrict the regions of the spacetime where such MOTS are allowed to be present.

Since conformal vector fields (and homotheties and isometries as particular cases) have very special deformation
tensors, the theorem above gives interesting information for spacetimes admitting such symmetries.

\begin{corollary}\label{thrstable} Let $\S$ be a
stable MOTS in a hypersurface $\Sigma$ of a spacetime $(M,g)$ which admits
a
conformal Killing vector $\vec{\xi}$,  $\mathcal{L}_{\xi} g_{\mu\nu} = 2 \phi g_{\mu\nu}$ (including
homotheties $\phi=C$, and isometries $\phi=0$).
\begin{itemize}
\item[(i)] If
$2 \vec{l} (\phi) + N (\kappa_l^2 + G_{\mu\nu} l^{\mu} l^{\nu} ) |_{S} \leq 0$ and
not identically zero, then $(\vec{\xi} \cdot \vec{l} ) |_S <0$.
\item[(ii)] If $S$ is strictly stable and
$2 \vec{l} (\phi) + N (\kappa_l^2 + G_{\mu\nu} l^{\mu} l^{\nu} ) |_{S} \leq 0$ then
$(\vec{\xi} \cdot \vec{l} ) |_S \leq 0$ and vanishes at one point only if it vanishes everywhere
\end{itemize}
\end{corollary}

{\bf Remark.} As before, the theorem is still true if all inequalities are reversed.

{\bf Remark.} In the case of homotheties and Killing vectors, the condition of the theorem
demands that $N_S W \leq 0$. Under NEC, this holds provided
$N_S \leq 0$, i.e. when $\vec{\xi}$  points below $\Sigma$ everywhere  on $S$ (where the term
``below'' includes also the tangential directions). For strictly stable $S$, the conclusion of the theorem
is that the homothety or the
Killing vector must lie above the null hyperplane defined
by the tangent space of $\S$
and the outer null normal $\vec{l}$ at each point $p \in S$.
If the
MOTS is only assumed to be stable, then the theorem requires the extra condition that
$\vec{\xi}$ points strictly below $\Sigma$ at some point with $W \neq 0$. However, the conclusion is also
stronger and forces $\vec{\xi}$ to lie strictly above the null hyperplane everywhere.
By changing the orientation of $\vec{\xi}$, it is clear that similar restrictions arise
when $\vec{\xi}$ is assumed to point {\it above} $\Sigma$. Figure 2
summarizes the allowed and forbidden regions for $\vec{\xi}$ in this case.

{\bf Proof.}
We only need to show that
$Z = 2 \vec{l} (\phi) + N (\kappa_l^2 + G_{\mu\nu} l^{\mu} l^{\nu} ) |_{S}$ for
conformal Killing vectors. This follows at once from (\ref{Z}) and
$a_{\mu\nu} = 2 \phi g_{\mu\nu}$ after using orthogonality of $\vec{e}_A$ and
$\vec{l}$. Notice in particular that $Z$ is the same for isometries and for
homotheties. 

$\hfill \Box$

\begin{figure}
\begin{center}
\includegraphics[width=9cm]{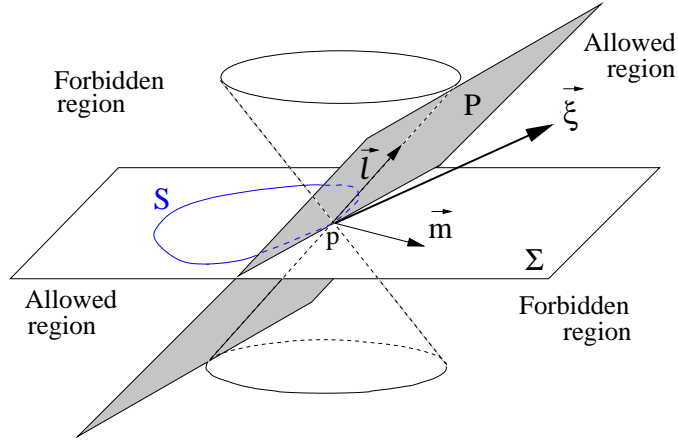}
\caption {The planes $T_p \Sigma$ and $P\equiv \{ q\in T_{p}M /
\left. Q \right|_{q}=0 \}$ divide the tangent space $T_{p}M$ in
four regions. By Corollary \ref{thrstable}, if $S$ is strictly
stable and $\vec{\xi}$ is a Killing vector or a homothety in a spacetime satisfying NEC
which points above $\Sigma$ everywhere, then
$\vec{\xi}$ cannot enter into the forbidden region at any point (and similarly, if
$\vec{\xi}$ points below $\Sigma$ everywhere).  The allowed region includes the plane $P$.
However, if there is a point with $W \neq 0$ where $\vec{\xi}$ is not tangent to 
$\Sigma$, then the result is also valid for stable MOTS and, moreover,
$P$ belongs to the forbidden region in this case.}
\end{center}
\end{figure}

This corollary has an interesting consequence in spacetime regions where there exists
a Killing vector or a homothety $\vec{\xi}$ which is causal everywhere.

\begin{corollary}
\label{shear}
Let a  spacetime $(M,g)$ satisfying NEC admit a
causal Killing vector or homothety $\vec{\xi}$ which is future
(past) directed everywhere on a stable MOTS $\S \subset \Sigma$.  Then,
\begin{itemize}
\item[(i)] The second fundamental form $\kappa^{l}_{AB}$ along $\vec{l}$
and $G_{\mu\nu}l^{\mu}l^{\nu}$ vanish identically on every point $p \in S$
where $\vec{\xi}|_p \neq 0$.
\item[(ii)] If $S$ is strictly stable, then $\vec{\xi} \propto \vec{l}$
everywhere.
\end{itemize}

\end{corollary}

{\bf Remark.} If we assume that there exists an open neighbourhood of $\S$ in $M$ 
where the Killing vector or homothety $\vec{\xi}$ is causal and future (past) 
directed everywhere then the conclusion (i) can be strengthened to say that
$\kappa^{l}_{AB}$ and $G_{\mu\nu}l^{\mu}l^{\nu}$ vanish identically on $S$.
The reason  is that such a $\vec{\xi}$ cannot vanish anywhere in this neighbourhood  (and 
consequently neither on $S$). For Killing vectors this result is proven
in Lemma 3.2 in \cite{BEM88} and a simple generalization shows that 
the same holds for homothetic Killing vectors.

{\bf Proof.}
We can assume, after reversing the sign of $\vec{\xi}$ if necessary,
that $\vec{\xi}$ is past directed, i.e. $N_S  \leq 0$.

Under NEC, $W$ is the sum of two non-negative terms, so in order to
prove (i) we only need to show that $W =0$ on points where $\vec{\xi} \neq 0$, i.e.
at points where $N_S <0$. Assume, on the contrary, that $W \neq 0$ and $N_S <0$
happen simultaneously at a point $p \in S$.
It follows that $N_S W \leq 0$ everywhere
and non-zero at $p$. Thus, we can apply statement (i) of Corollary \ref{thrstable} to conclude
$Q <0$ everywhere. Hence $N_S Q \geq 0$ and not identically zero on $S$. Recalling the decomposition
$\vec{\xi} = N_S \vec{l} + Q \vec{m} + \vec{Y}^{\parallel}$, the square  norm of this vector is
\begin{eqnarray}
\label{normsquare}
\left ( \vec{\xi} \cdot \vec{\xi} \, \right ) = 2N_{\S}Q+Q^2+
\left ( \vec{Y}^{\parallel}\cdot\vec{Y}^{\parallel}  \right ).
\end{eqnarray}
This is the sum of non-negative terms, the
first one not identically zero. This contradicts the condition of $\vec{\xi}$ being causal.

To prove the second statement, we notice that
point (ii) in Corollary \ref{thrstable} implies $Q \leq 0$, and hence
$N_S Q \geq 0$. The only possibility how (\ref{normsquare}) can be negative or zero, is
$Q= \vec{Y}^{\parallel} =0$, i.e. $\vec{\xi} \propto \vec{l}$. $\hfill \Box$

This corollary extends Theorem 2 in \cite{MS03} to the case of
stable MOTS and implies, for instance, that any
strictly stable MOTS in a plane wave spacetime (which by definition
admits a null and nowhere zero Killing vector field $\vec{\xi}$)
must be aligned with the direction of propagation of the wave (in the
sense that $\vec{\xi}$ must be one of the null normals to the surface).
It also implies that any spacetime admitting a causal and future
directed Killing vector (or homothety) whose energy-momentum tensor
does not admit a null eigenvector (e.g. a perfect fluid) cannot contain
any stable MOTS.

The results above hold for stable or strictly stable MOTS. Among such surfaces,
marginally trapped surfaces are of special interest. Our next result restricts (and in some cases
forbids) the existence of such surfaces in spacetimes admitting Killing vectors, homotheties
or conformal Killings.

\begin{thr}
\label{thrfirstvariation}
Let $\S$ be a stable MOTS
in a spacelike hypersurface $\Sigma$ of a spacetime $(M,g)$ which satisfies NEC and
admits a conformal Killing vector
$\vec{\xi}$ with conformal factor $\phi \geq 0$ (including homotheties with $C\geq0$ and Killing vectors).
Suppose furthermore that
either (i) $( 2\vec{l}(\phi)+NW  ) |_S \not \equiv 0$ or (ii)
$S$ is strictly stable and $( \vec{\xi} \cdot \vec{l} ) |_S \not \equiv 0$. Then
the following holds.
\begin{itemize}
\item[(a)]
If $ 2\vec{l}(\phi)+NW  |_S \leq 0$  then
$S$ cannot be a marginally future trapped surface, unless $\vec{H} \equiv 0$. The latter
case is excluded if $\phi|_S \not \equiv 0$.
\item[(b)] If $ 2\vec{l}(\phi)+NW |_S \geq 0$ then
$S$ cannot be a marginally past trapped surface, unless $\vec{H} \equiv 0$. The latter
case is excluded if $\phi|_S \not \equiv 0$.
\end{itemize}
\end{thr}

{\bf Remark.} The statement obtained from this one by reversing all the inequalities is also true.
This is a direct consequence of the freedom in changing
$\vec{\xi} \rightarrow - \vec{\xi}$.

{\bf Proof.} We will only prove case (a). The argument for case (b) is similar.
The idea is taken from \cite{MS03} and consists of performing 
a variation of $S$ along the conformal Killing vector and evaluate
the change of area in order to get a contradiction if $S$ is marginally future trapped. 
The difference is that here do not make any a priori assumption 
on the causal character for $\vec{\xi}$. Corollary \ref{thrstable} 
provides us with sufficient information for the argument to go through.

As before, let $S^{\prime}_t$ be the collection of surfaces obtained by displacing $S$
with the local diffeomorphism generated by $\vec{\xi}$ a parametric amount $t$.
We denote by $|S^{\prime}_t|$ their corresponding areas.
The first variation of area (see e.g. \cite{MS03}) gives
\begin{equation}\label{variationofareaMOTS}
\left.\frac{d |S^{\prime}_t|}{dt}\right|_{t=0}=-\frac12\int_{S}\theta_{k} \left ( \vec{\xi} \cdot \vec{l} \, \right )
\eta_{\S},
\end{equation}
where $\eta_{\S}$ is the volume form of $\S$ and we have used $\vec{H} = -
\frac{1}{2} \theta_k \vec{l}$.  Now, since $2\vec{l}(\phi)+NW  |_S \leq 0$, and
furthermore either hypothesis (i) or (ii) holds,  Corollary
\ref{thrstable} implies that $( \vec{\xi} \cdot \vec{l} ) |_S < 0$.

On the other hand, being $\vec{\xi}$ a conformal Killing vector,
the induced metric on $S^{\prime}_t$ is related to the metric on $S$ by
conformal rescaling. A simple calculation gives (see e.g. \cite{MS03})
\begin{equation}\label{variationofareaconformalKilling}
\left.\frac{d |S^{\prime}_t|}{dt}\right|_{t=0}=2\int_S \phi\eta_S,
\end{equation}
This quantity is non-negative due to $\phi\geq 0$ and not identically zero if $\phi \neq 0 $ somewhere.
Combining (\ref{variationofareaMOTS}) and (\ref{variationofareaconformalKilling}) we conclude that
if $\theta_k \leq 0 $ (i.e. $S$ is marginally future trapped) then necessarily $\theta_k$ vanishes identically
(and so does $\vec{H}$). Furthermore, if $\phi |_S$ is non-zero somewhere, then $\theta_k$ must necessarily
be positive somewhere, and $S$ cannot be future marginally trapped. $\hfill \Box$

\subsection{An application: No stable MOTSs in FLRW}
\label{sectionFLRW}

In this subsection we apply Corollary \ref{thrstable} to show that 
a large subclass of  Friedmann-Lema\^itre-Robertson-Walker (FLRW)
spacetimes do not admit stable MOTS on {\it any} spacelike hypersurface.
Obtaining the corresponding results for round spheres only requires
a straightforward calculation, and is therefore simple. The power 
of the method is that it provides a general result involving no
assumption on the geometry of the MOTS or on the spacelike hypersurface
where it is embedded. The only requirement is that the scale factor
and its time derivative satisfy certain inequalities. This includes,
for instance all FLRW cosmologies satisfying NEC and with 
accelerated expansion, as we shall see in Corollary \ref{corollaryFLRW} below.

Recall that the FLRW metric is 
\begin{eqnarray*}
g_{FLRW}=-dt^2+a^{2}(t)\left[ dr^2+\chi^{2}(r;k)d\Omega^{2} \right],
\end{eqnarray*}
where $a(t)>0$ is the scale factor
and $\chi(r;k)= \{\sin{r}, r, \sinh r \}$ for $k=\{1,0,-1\}$ respectively.
The Einstein tensor of this metric is of perfect fluid type (see e.g. \cite{Wald84})
and reads
\begin{eqnarray}
G_{\mu\nu} = (\rho + p) u_{\mu} u_{\nu} + p g_{\mu\nu}, \quad \vec{u} = \partial_t, \quad 
\rho = \frac{3(\dot{a}^2(t)+k)}{a^2(t)}, \quad
\rho + p =  2\left ( \frac{\dot{a}^2(t)+k}{a^2(t)}-
\frac{\ddot{a}(t)}{a(t)} \right)
\label{EM_FLRW}
\end{eqnarray}
where dot stands for derivative with respect to $t$. 
\begin{thr}\label{thrFRW} 
There exists no stable MOTS in any spacelike 
hypersurface of a FLRW spacetime $(M,g_{FLRW})$ satisfying 
\begin{equation}\label{conditionFLRW}
\frac{{\dot{a}}^{2}(t)+k}{a(t)} >0 , \quad
-\frac{\dot{a}^{2}(t)+k}{a(t)}\leq
\ddot{a}(t)\leq \frac{{\dot{a}}^{2}(t)+k}{a(t)}.
\end{equation} 
\end{thr}

{\bf Remark.} In terms of the energy-momentum contents of the spacetime, these three conditions read,
respectively, $\rho \geq 0$, $\rho \geq 3 p$ and $\rho + p \geq 0$. As an example,
in the absence of a
cosmological constant they are satisfied
as soon as energy conditions are imposed and the pressure 
is not too large (e.g. for the matter and radiation dominated eras). The class of
FLRW satisfying (\ref{conditionFLRW}) is clearly very large. We also remark that 
Theorem \ref{thrFRW} agrees with the fact \cite{S97} that 
the causal character of the hypersurface which separates the trapped from the
non-trapped {\it spheres} in FLRW spacetimes depends precisely on the quatity $\rho^2(\rho+p)(\rho-3p)$.

{\bf Proof.} The FLRW spacetime admits a conformal Killing vector 
$\vec{\xi}=a(t) \vec{u}$ with conformal factor $\phi=\dot{a}(t)$. 
Since this vector is timelike and future directed, it follows 
that  $( \vec{\xi}\cdot\vec{l} )|_S<0$
for any spacelike surface $S$ embedded in a spacelike hypersurface $\Sigma$.
If we can show that
$\left. 2\vec{l}(\phi)+N(\kappa_l^2+G_{\mu\nu}l^{\mu}l^{\nu})\right|_S \geq 0$,
and non-identically zero for any $S$, then point (i) in Corollary 
(\ref{thrstable}) 
implies that $S$ cannot be a stable MOTS, thus proving the
result. The proof therefore relies on finding 
conditions on the scale factor which imply the validity of this inequality on any $S$.
First of all, we notice that the second fundamental form
$\kappa^l_{AB}$ can be made as small as desired on 
a suitably chosen $S$. Thus, the inequality that needs to be satisfied is
\begin{equation}\label{conditioncorollary2}
\left. 2\vec{l}(\phi)+N G_{\mu\nu}l^{\mu}l^{\nu}\right|_S \geq 0,
\end{equation}
and positive somewhere. In order to evaluate this expression
recall that $\vec{u} = a^{-1} \vec{\xi} = a(t)^{-1} N \vec{n} + a(t)^{-1} \vec{Y}$.  
Let us write
$\vec{Y} = Y \vec{e}$, where $\vec{e}$ is unit and let $\alpha$ be the hyperbolic angle of $\vec{u}$ in the 
basis $\{\vec{n}, \vec{e} \}$, i.e. $\vec{u} = \cosh \alpha \, \vec{n} + \sinh \alpha \vec{e}$. It follows immediately
that $N = a(t) \cosh \alpha$ and $Y = a(t) \sinh \alpha$. Furthermore, multiplying $\vec{u}$ by the normal
vector to the surface we find $( \vec{u} \cdot \vec{m} ) = \cos \varphi \sinh \alpha$, where
$\varphi$ is the angle between $\vec{m}$ and $\vec{e}$. With this notation, let us calculate 
the null vector $\vec{l}$. Writing $\vec{l} = A \vec{u}  + \vec{b}$, with
$\vec{b}$ orthogonal to $\vec{u}$, it follows $(\vec{b} \cdot \vec{b}) = A^2$ from the condition of
$\vec{l}$ being  null. On the other hand we have the decomposition $
A \vec{u} + \vec{b} = \vec{l} = \vec{n} + \vec{m}$. Multiplying by $\vec{u}$ we immediately get
$A =  \cosh \alpha - \cos \varphi \sinh \alpha$, and since $\phi = \dot{a}(t)$ only depends on $t$
\begin{equation}\label{l(phi)FRW}
\vec{l}(\phi)=\left( \cosh{\alpha}-\cos{\varphi}\sinh{\alpha} \right)\ddot{a}(t).
\end{equation}
The following expression for  $G_{\mu\nu}l^{\mu}l^{\nu}$ follows directly from $\vec{l} = A \vec{u} + \vec{b}$ and
(\ref{EM_FLRW}),
\begin{eqnarray}
\label{gll}
G_{\mu\nu}l^{\mu}l^{\nu}=A^{2}(\rho+ p ) = 
2\left( \cosh{\alpha}-\cos{\varphi}\sinh{\alpha} \right)^2 
\left ( \frac{\dot{a}^2(t)+k}{a(t)}-\frac{\ddot{a}(t)}{a(t)} \right).
\end{eqnarray}
Inserting  (\ref{l(phi)FRW}) and (\ref{gll}) into (\ref{conditioncorollary2})
and dividing by $A \cosh \alpha$ (which is positive) we find the equivalent condition
\begin{equation}\label{conditioncorollary2c} 
\left (
\frac{1}{\cosh{\alpha}\left(\cosh{\alpha}-\cos{\varphi}\sinh{\alpha}\right)}-1
\right )\ddot{a}(t)+\frac{\dot{a}^{2}(t)+k}{a(t)} \geq 0,
\end{equation}
and non-zero somewhere. The dependence on $S$ only arises 
through the function
$f(\alpha,\varphi) = \cosh \alpha ( \cosh \alpha - \cos \varphi \sinh \alpha)$.
Rewriting this as $f = 1/2 [ 1 + \cosh (2 \alpha) - \cos \varphi \sinh (2\alpha) ]$
it is immediate to show that $f$ takes all values in $(1/2, +\infty)$. Hence
\begin{eqnarray*}
-1 < \left (
\frac{1}{\cosh{\alpha}\left(\cosh{\alpha}-\cos{\varphi}\sinh{\alpha}\right)}-1 \right ) < 1. 
\end{eqnarray*}
In order to satisfy (\ref{conditioncorollary2c}) on all this range, it is necessary
and sufficient that the two inequalities in (\ref{conditionFLRW}) are satisfied
$\hfill \square$

The following Corollary gives a particularly interesting case 
where all the conditions of Theorem \ref{thrFRW}
are satisfied.

\begin{corollary}
\label{corollaryFLRW}
Consider a FLRW spacetime $(M,g_{FLRW})$ satisfying NEC. If $\ddot{a}(t)>0$, then there 
exists no stable MOTS in any spacelike hypersurface of $(M,g_{FLRW})$
\end{corollary}

{\bf Proof.} 
The null energy condition gives  $0 \leq \rho+p =2\left ( 
\frac{\dot{a}^2(t)+k}{a^2(t)} -\frac{\ddot{a}(t)}{a(t)} \right)$. This implies the first
two inequalities in (\ref{conditionFLRW}) if $\ddot{a}>0$. The remanining condition
$- \frac{\dot{a}^2(t)+k}{a(t)} \leq \ddot{a}$ is also obviously satisfied provided $\ddot{a} >0$.
$\hfill \square$

\subsection{A consequence of the geometric construction of $\{S_t\}$}
\label{sectiongeometric}

We have emphasized at the beginning of this section that the restrictions obtained directly
by the geometric procedure
of moving $S$ along $\vec{\xi}$ and then back to $\Sigma$  are intuitively
clear but typically weaker than those obtained by using elliptic theory results.
There are some cases, however, where
the reverse  actually holds, and the geometric construction provides stronger results. 
We will present
one of these cases in this subsection.

Corollary \ref{thrstable} gives restrictions on $(\vec{\xi}\cdot \vec{l}) |_S$ for Killing vectors
and homotheties in spacetimes satisfying NEC, provided $\vec{\xi}$ is future or past
directed everywhere. However, when $W$ vanishes identically, the result only gives useful information in the
strictly stable case. The reason is that
$W \equiv 0$ implies $L_m Q \equiv 0$ and, for marginally stable surfaces (i.e. $\lambda=0$), the maximum
principle is not strong enough to conclude that $Q$ must have a sign. There is at least one case where marginally
stable surfaces play an important role, namely after a jump in the outermost MOTS in a 3+1 foliation
of the spacetime (see \cite{AMMS09} for details). As we will see next, the geometric construction does
give restrictions in this case even when $W$ vanishes identically. Let us start by recalling
the definition of {\bf outermost} MOTS.

Let $\Sigma$ be a spacelike hypersurface whose boundary consists of the union of
two disjoint sets $\partial\Sigma=\partial^{+}\Sigma\cup\partial^{-}\Sigma$.
We take $\Sigma$
to be disjoint to its boundaries and assume that $\Sigma$ has compact closure.
Endow $\partial^{+} \Sigma$  with an outer normal pointing outside $\Sigma$ and
$\partial^{-} \Sigma$  with an outer normal pointing inside $\Sigma$. Assume that the outer boundary
$\partial^{+} \Sigma$ is outer untrapped $\theta_{l}^{+}>0$ and that
the inner boundary $\partial^{-} \Sigma$ is weakly outer trapped $\theta_{l}^{-}\leq 0 $.
Under these conditions,
Theorem 7.3 of  \cite{AM07} asserts that there always exists a unique
outermost MOTS $S \subset \Sigma
\cup \partial^-\Sigma$ homologous
to $\partial^{+} \Sigma$ (i.e. such that together with the outer boundary it bounds an open domain
${\cal V}$). Outermost means that no weakly outer trapped surface contained in $\Sigma \cup 
\partial^- \Sigma$ and homologous
to the outer boundary can intersect ${\cal V}$.
Obviously, the outermost MOTS is locally outermost and hence necessarily stable.
When requiring a surface $S$ to be outermost, we will implicitly assume all the above conditions
on $\Sigma$.

We can now state the following result

\begin{thr}\label{thrkilling} Consider a spacetime $(M,g)$ possessing a
Killing vector or a homothety $\vec{\xi}$ and satisfying
NEC. Let $S$ be the outermost MOTS on a spacelike hypersurface
$\Sigma$ defined locally by a level function $T=0$ with $T>0$ to the future of
$\Sigma$. If $\vec{\xi}(T) \leq 0 $ on some spacetime neighbourhood
of $S$, then $(\vec{\xi} \cdot \vec{l} )\leq 0$ everywhere on $\S$.
\end{thr}

{\bf Remark.} As usual, the theorem still holds if all the inequalities are reversed.

{\bf Remark.} The simplest way to ensure that $\vec{\xi}(T)\leq 0 $ on some neighbourhood
of $S$ is by imposing a condition merely on $\S$, namely $(\vec{\xi} \cdot \vec{n} ) |_S > 0 $, 
because then $\vec{\xi}$ lies
strictly below $\Sigma$  on $S$ and this property is obviously preserved sufficiently near $S$ (i.e.
$\vec{\xi}$ points strictly below the level set of $T$ on a sufficiently small spacetime neighbourhood of $S$).

{\bf Proof.} The idea is to use the geometric procedure described above to construct $\{S_t\}$
and use the fact that $S$ is outermost to conclude that $\{S_t\}$ ($t>0$) cannot have points
outside $S$. Here we move $S$ a small but finite amount $t$, in contrast
to the elliptic results before, which only involved infinitesimal displacements. We want to have information
on the sign of the outer expansion of $S_t$ in order to make sure that a weakly outer trapped surface forms.
The first part of the displacement is along $\vec{\xi}$ and gives $S^{\prime}_t$.
Let us first see that all these surfaces are MOTS. For Killing vectors,
this follows at once from symmetry arguments. For homotheties ($\mathcal{L}_{\xi} \, g_{\alpha\beta} =
2 C g_{\alpha\beta}$) we have the identity
\begin{equation}\label{noindependentterm}
\delta_{\xi}\theta_{l}= \left ( -\frac{1}{2}
k^{\alpha}\mathcal{L}_{{\xi}}
{l_t}^{\prime}_{\alpha} 
-2C
\right )\theta_{l},
\end{equation}
which follows directly from (\ref{xithetau}) with $\vec{\eta} = \vec{l}$ after using
$l^{\mu} \mathcal{L}_{\xi} \, {l_t}^{\prime}_{\mu}  
 = \frac{1}{2} a_{\mu\nu} l^{\mu} l^{\nu} = 0$, see
(\ref{null}). Expression (\ref{noindependentterm}) holds
for each one of the surfaces $\{S_t\}$, independently of them being MOTS or not.
Since this variation vanishes on MOTS and the starting surface $S$ has this property,
it follows that each surface $S^{\prime}_t$ ($t>0$) is also a MOTS. Moving back to $\Sigma$ along the null
hypersurface introduces, via the Raychaudhuri equation, a non-positive term in the outer null expansion,
provided the motion is to the future. Hence, $S_t$ for small but finite $t>0$ is a weakly outer trapped surface
provided $\vec{\xi}$ moves to the past of $\Sigma$. This is ensured if $\vec{\xi} (T) \leq 0$ near $S$, because
$T$ cannot become positive for small enough $t$. On the other hand, since a point $p \in S$ moves initially
along the vector field $\nu = \vec{\xi} - N_S \vec{l} = Q \vec{m} + \vec{Y}^{\parallel}$, where 
$Q=(\vec{\xi}\cdot\vec{l})$ as usual, it follows that
$Q>0$ somewhere  implies (for small enough $t$) that
the weakly outer trapped surface $S_t$ has a portion lying strictly to the outside of $S$, which
is a contradiction to $S$ being outermost. Hence $Q\leq 0$ everywhere and the theorem is proven. $\hfill \Box$

It should be remarked that the assumption of $\vec{\xi}$ being a Killing vector or a homothety is important
for this result. Trying to generalize it for instance to conformal Killings fails in general because
then the right hand side of
equation (\ref{noindependentterm}) has an additional term $2\vec{l}(\phi)$, not proportional to
$\theta_l$. This means that moving a MOTS along
a conformal Killing does not lead to another MOTS in general. The method can however, still give useful information
if $\vec{l} (\phi)$ has the appropriate sign, so that $S^{\prime}_t$ is in fact weakly
outer trapped. We omit the details.

So far, all the results we have obtained require that the quantity $L_{m}Q$ does not change
sign on the MOTS $\S$. In the next section we will relax this condition.

\section{Results regardless of the sign of $L_m Q$}
\label{sectionnonelliptic}

When $L_m Q$ changes sign on $\S$, the elliptic methods exploited in the previous section
loose their power. Moreover, for sufficiently small $t$, the surface $\{\S_t\}$
defined by the geometric construction above necessarily fails to be weakly outer trapped. 
Thus, obtaining restrictions in this case becomes a much harder problem.

However, for locally outermost MOTS $S$,
an interesting situation arises when $\S_t$ lies partially outside $S$ and 
happens to be weakly outer trapped in that exterior region.
More precisely, if a connected component of the subset of $S_t$ which lies outside 
$S$ turns out to have non-positive outer null expansion, then using a smoothing result 
by Kriele and Hayward \cite{KH97}, we will be able to construct a new weakly outer trapped surface outside
$S$, thus leading to a contradiction with the fact that $S$ is locally 
outermost (or else giving restrictions on the
generator $\vec{\xi}\, $).

The result by Kriele and Hayward states, in rough terms, 
that given two surfaces which intersect on a curve, a new smooth surface
can be constructed lying outside the previous ones in such a way that the
outer null expansion does not increase in the process. The precise statement is as follows.

\begin{lema}\label{lemasmoothness}
Let $S_{1},S_{2}\subset \Sigma$ be
smooth two-sided surfaces which intersect transversely on a smooth
curve $\gamma$.  Assume it is possible to choose one connected
component of each set $S_{1}\setminus
\gamma$ and $S_{2}\setminus\gamma$, say $S^{+}$ and $S^{-}$ respectively,
such that the outer normal $\vec{m}_{+}$ of $S^{+}$
and the vector $\vec{e}_{-}$ orthogonal to $\gamma$,
tangent to $\S^{-}$ and pointing towards $S^{-}$ satisfy ${m_{+}}_{\mu}{e_{-}}^{\mu}\geq 0$ everywhere 
on $\gamma$. Then, for any neighbourhood $V$ of $\gamma$ in $\Sigma$
there
exists a smooth surface $\tilde{S}$ and a continuous and piecewise
smooth bijection $\Phi\colon S^{+}\cup S^{-}\cup \gamma\rightarrow
\tilde{S}$ such that
\begin{enumerate}
\item $\Phi(p)=p$,  $\forall p\in\left( S^{+} \cup S^{-}\right)\setminus
V$
\item $\left.\tilde{\theta}_{l}\right|_{\Phi(p)}\leq
\left.{\theta}^{\pm}_{l}\right|_{p}$ $\forall p\in
S^{\pm}$, where $\tilde{\theta}_{l}$ is the null expansion of $\tilde{\S}$
and ${\theta}^{\pm}$ is the null expansion of $\S^{\pm}$.
\end{enumerate}
Moreover $\tilde{S}$ lies in the connected component
of $V\setminus\left( S^{+}\cup S^{-}\cup\gamma \right)$ into which
$\vec{m}_{+}$ points.
\end{lema}

This result will allow us to adapt the arguments above without having to assume
that  $L_m Q$ has a constant sign on $S$. The argument will be again by contradiction,
i.e. we will assume a locally outermost MOTS $S$ and, under suitable 
circumstances, we will be able to find a new weakly outer trapped
surface lying outside $S$. Since the conditions are much weaker than in the previous section, 
the conclusion is also weaker. It is, however, fully general in the sense that it holds for 
any vector field $\vec{\xi}$ on $S$. Recall that $Z$ is defined in equation (\ref{Z}).

\begin{thr}\label{thrnonelliptic} 
Let $S$ be a locally outermost MOTS in a
spacelike hypersurface $\Sigma$ of a spacetime $(M,g)$. Denote by $\U_0$
a connected component of
the set $\{p\in\S ; (\vec{\xi}\cdot\vec{l}) |_{p}>0 \}$. Assume
$U_0\neq \emptyset$ and that its boundary $\gamma\equiv\partial U_0$ is either empty, or it satisfies
that the function $(\vec{\xi}\cdot\vec{l})$ has a non-zero gradient everywhere on $\gamma$, i.e.
$d(\vec{\xi}\cdot\vec{l}) |_{\gamma} \neq 0$.

Then, there exists $p\in\overline{U_0}$ such that $\left.Z\right|_p\geq 0$.
\end{thr}

{\bf Proof.}  As mentioned, we will use a contradiction argument. Let us therefore assume that
\begin{equation}\label{conditionsubset}
Z|_p <0, \quad \forall p \in \overline{\U_0}.
\end{equation}
The aim is to construct a weakly outer trapped surface near $S$ and outside of it. This will contradict
the condition of $S$ being locally outermost.

First of all we observe that $Z$ cannot be negative everywhere on $S$, because
otherwise Theorem \ref{TrhAnyXi}
(recall that outermost MOTS are
always stable) would imply $Q\equiv(\vec{\xi}\cdot\vec{l}) <0$
everywhere and $U_0$ would be
empty against hypothesis. Consequently, under (\ref{conditionsubset}),
$U_0$ cannot coincide with $S$ and $\gamma\equiv \partial \U_0 \neq \emptyset$. Since
$\left.Q\right|_{\gamma}=0$ and, by assumption,
$\left. dQ \right|_{\gamma}\neq 0$ it follows that
$\gamma$  is a smooth embedded curve.
Taking $\mu$ to be a local coordinate on $\gamma$, it is clear that
$\{\mu,Q\}$ are coordinates of a neighbourhood of $\gamma$ in $S$.
We will coordinate a small enough neighbourhood of $\gamma$ in $\Sigma$
by Gaussian coordinates $\{ u,\mu,Q\}$ such that $u=0$ on $S$ and $u>0$
on its exterior.

By moving $S$ along $\vec{\xi}$ a finite but small parametric amount $t$
and back to $\Sigma$ with the outer null geodesics, as described in
Section \ref{sectionbasics}, we construct a family of surfaces $\{ S_t \}$.
The curve that each point $p \in S$ describes via this construction has tangent
vector $\nu = Q\vec{m}+\vec{Y}^{\parallel} |_{\S}$
on $S$. In a small neighbourhood of $\gamma$, the normal component of this vector, i.e.
$Q \vec{m}$, is smooth and only vanishes on $\gamma$. This implies that
for small enough $t$, $S_t$ are graphs over $S$ near $\gamma$. We will always
work on this neighbourhood, or suitable restrictions thereof.
In the Gaussian coordinates above this graph is of the form $\{ u=\hat{u}(\mu,Q,t),\mu,Q \}$.
Since the normal unit vector to $S$ is simply
$\vec{m} = \partial_u$ in these coordinates and the normal component of $\nu$ is $Q \vec{m}$, 
the graph function $\hat{u}$ has the 
following Taylor  expansion 
\begin{equation}\label{graph}
\hat{u}(\mu,Q,t)=Qt+O(t^2).
\end{equation}
Our next aim is to use this expansion to conclude that 
the intersection of $S$ and $S_t$ near
$\gamma$ is an embedded curve $\gamma_t$ for all small enough $t$. To do that
we will apply the implicit function theorem to the equation $\hat{u}=0$. It is useful
to introduce a new function
$v(\mu,Q,t)=\frac{\hat{u}(\mu,Q,t)}{t}$, which is still smooth (thanks to 
(\ref{graph})) and vanishes
at $t=0$ only on the curve $\gamma$. Moreover, its derivative with respect to
$Q$
is nowhere zero on $\gamma$, in fact
$\left.\frac{\partial
v}{\partial Q}\right|_{(\mu,0,0)}=1$ for all $\mu$.
The implicit function theorem implies that there
exist a unique function $Q=\varphi(\mu,t)$ which solves the equation
$v(\mu,Q,t)=0$, for small enough $t$. Obviously, this function is also the
unique solution near $\gamma$
of $\hat{u}(\mu,Q,t)=0$ for $t>0$. Consequently, 
the intersection of $S$ and $S_t$ ($t>0$) 
lying in the
neighbourhood of $\gamma$ where we are working on is an embedded curve $\gamma_t$.
This curve divides $S_t$ into two connected components (because $\gamma$ does).
Let us denote by $S_t^+$ the connected component of $S_t$ which has
$v (\mu,Q,t) >0$ near
$\gamma$ (i.e. that lies in the exterior of $S$ near $\gamma$). This connected component
in fact lies fully outside of $S$, not just in a neighborhood of $\gamma$, as we see next.
First of all, recall that $\gamma$ is the boundary of a connected set $\U_0$ where $Q$
is strictly positive. We have just seen that $\gamma_t$ is a continuous 
deformation of $\gamma$. Let us denote by $U_t$ the domain 
obtained by deforming $U_0$  when the boundary moves from $\gamma$ to $\gamma_t$ (See Fig.3).
It is obvious that $S_t^+$ is obtained by moving $U_t$ first along $\vec{\xi}$
an amount $t$ and the back to $\Sigma$ by null hypersurfaces. The closed subset of $U_t$
lying outside the tubular neighbourhood
where we applied the implicit function theorem is, by construction
a proper subset of $U _0$.
Consequently, on this closed set $Q$ is uniformly bounded below
by a positive constant. Given that $Q$ is the first order order term of the normal variation,
all these points move outside of $S$. This proves that $S^+_t$ is fully outside $S$ for
sufficiently small $t$. Incidentally this also shows that $S^+_t$ is a graph over $U_t$.

\begin{figure}
\begin{center}
\includegraphics[width=10cm]{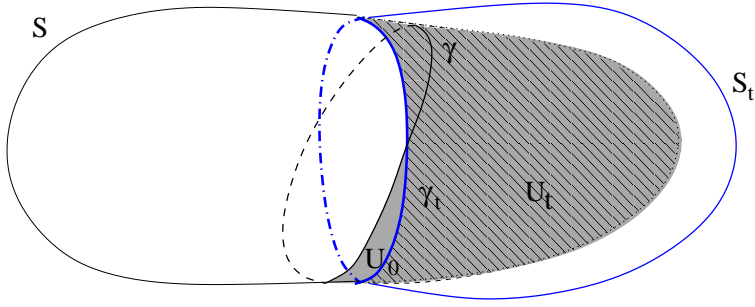}
\caption {The figure represents both intersecting surfaces $S$ and $S_t$ together with the curves 
$\gamma$ and $\gamma_t$. The shaded region corresponds to $U_0$ and the region with lines to $U_{t}$.}
\end{center}
\end{figure}

The next aim is to show that the outer null expansion
of $\S_t$ is non-positive everywhere on $\S^{+}_t$.
To that aim, we will prove that, for small enough $t$, $Z$ is strictly negative 
everywhere on $U_t$. Since $Z$ is the first order term in the variation of $\theta_l$, this
implies that the outer null expansion of $S^+_t$ satisfies $\theta^+_{l, t} <0$ for $t>0$
small enough.

By assumption (\ref{conditionsubset}), $Z$ is strictly negative on $U_0$. Therefore, this 
quantity is automatically negative in the portion  of $U_t$ lying in $U_0$ (in particular,
outside the tubular neighbourhood
where we applied the the implicit function theorem). The only difficulty comes from the fact that
$\gamma_t$ may move outside $U_0$ at some points and we only have information on the sign 
of $Z$ on $\overline{U_0}$. To address this issue, we first notice that $Q$ defines 
a distance function to $\gamma$ (because $Q$ vanishes on $\gamma$ and its gradient is nowhere zero).
Consequently, the fact that $Z$ is strictly negative on $\gamma$ (by assumption
(\ref{conditionsubset})) and that this curve is compact imply that there exists a $\delta>0$
such that, inside the tubular neighbourhood of $\gamma$, $|Q|<\delta$  implies $Z >0$.
Moreover, the function $Q = \phi(\mu,t)$, which defines $\gamma_t$, is such that 
it vanishes at $t=0$ and depends smoothly in $t$. Since $\mu$ takes values on a compact set, it follows
that for each $\delta^{\prime}>0$, there exists an $\epsilon(\delta^{\prime})>0$, independent of
$\mu$ such that
$|t| < \epsilon(\delta^{\prime})$ implies $|Q| = |\phi(\mu,t)| < \delta^{\prime}$. By taking $\delta^{\prime} =
\delta$, it follows that, for $|t| < \epsilon (\delta)$, $U_t$ is contained in a
$\delta$-neighbourhood of $U_0$ (with respect to the distance function $Q$) and consequently $Z<0$ 
on this set, as claimed. We restrict to $0 < t < \epsilon(\delta)$ from now on.

Summarizing, so far we have shown that $\S_{t}^{+}$ lies fully outside $\S$ and
has $\theta_{l,t}^{+}<0$. The final task is to use 
Lemma \ref{lemasmoothness} to construct 
a weakly outer trapped surface strictly outside $\S$.
Indeed, the curve $\gamma_{t}$ divides the locally
outermost MOTS $\S$ in two connected components. Since $S$ is locally outermost, 
there is a two-sided neighbourhood of $S$ in $\Sigma$. Following the notation
in Section \ref{sectionbasics}, we call $\mathfrak{D}$ the interior part of
this two-sided neighbourhood. Denote by $S^-_t$ the complementary of $U_t$
in $S$. By construction, $\S_{t}^+\cup\gamma_{t}\cup\S^{-}_t$ bounds a domain which
contains $\mathfrak{D}$. Now, let $\vec{e}_{-}$ be the vector normal
to $\gamma_{t}$ and tangent to $S^{-}_t$ that points to the interior
of $S^{-}$ and let $\vec{m}_{+}$ be the vector normal to $S_{t}^{+}$ which
points to the exterior of $S_{t}^+$. Since $S^{+}_{t}$ lies outside $\S$ 
and it is a graph on $S$ near $\gamma_t$, it follows immediately that
${m_{+}}_{\mu}{e_{-}}^{\mu}\geq 0$ holds
everywhere on $\gamma_{t}$.  Therefore, Lemma
\ref{lemasmoothness} guarantees that there exists a weakly outer
trapped surface $\tilde{S}$ lying outside $\S$, leading to a contradiction.
$\hfill \square$.

\vskip 2mm

{\bf Remark.} As always, this theorem also holds if all the inequalities are reversed. 
Note that in this case $U_0$ is defined to be a connected component of the set 
$\{p\in\S; (\vec{\xi}\cdot\vec{l}) |_p<0 \}$. For the proof simply take $t<0$
instead of $t>0$ (or equivalently move along $-\vec{\xi}$ instead of $\vec{ \xi}$).

\vskip 2mm

Similarly as in the previous section, this theorem can be particularized to the case
of conformal Killing vectors, as follows

\begin{corollary}\label{corollarynonelliptic}
Under the assumptions of Theorem \ref{thrnonelliptic}, suppose that $\
\vec{\xi}$ is a conformal Killing vector with conformal factor $\phi$
(including homotheties $\phi=C$ and isometries $\phi=0$).

Then, there exists $p\in\overline{U_0}$ such that 
$2\vec{l}(\phi)+N_{S}(\kappa_{l}^{2}+G_{\mu\nu}l^{\mu}l^{\nu}) |_p\geq 0$
\end{corollary}

If the conformal Killing is in fact a homothety or a Killing vector 
and it is causal everywhere, the result can be strengthened  considerably. The
next result extends in a suitable sense Corollary 
\ref{shear} to the cases when the generator is not assumed to be either
future or past everywhere. Since its proof requires an extra
ingredient we write it down as a theorem

\begin{thr}\label{shear2}
In a spacetime $(M,g)$ satisfying NEC and admitting a 
Killing vector or homothety  $\vec{\xi}$, consider a locally outermost MOTS 
$\S$ in a spacelike hypersurface $\Sigma$. Assume that $\vec{\xi}$ is causal on $S$ and that
$W=\kappa_{l}^2+G_{\mu\nu}l^{\mu}l^{\nu}\neq 0$  everywhere. Define 
$U\equiv\{p\in\S ; (\vec{\xi}\cdot\vec{l})|_{p}>0\}$ and assume that this set
is neither empty nor covers all of $S$. Then, on each connected component $U_{i}$ of $U$ there exist a 
point $p\in\partial U_{i}$ with $d(\vec{\xi}\cdot\vec{l})|_{p}=0$
\end{thr}

{\bf Remark.} The same conclusion holds on the boundary of each connected components of the set $\{
p\in\S ; (\vec{\xi}\cdot\vec{l})|_{p}<0\}$. This is obvious since $\vec{\xi}$ can be changed
to $-\vec{\xi}$.

{\bf Remark.} The case $\partial U=\emptyset$, excluded by assumption in this theorem, can only occur
if $\vec{\xi}$  is future or past everywhere on $S$. Hence, this case is already
included in  Corollary \ref{shear}.

{\bf Proof.} We first show that on any point in $U$ we have $N_S < 0$, which 
has as an immediate consequence that $N_S \leq 0$ on any point in $\overline{U}$. 
The former statement is a consequence of the decomposition
$\vec{\xi}=N\vec{l}+Q\vec{m}+\vec{Y}^{\parallel}$, where 
$Q=(\vec{\xi}\cdot\vec{l})$. The condition that $\vec{\xi}$ is causal then implies
$( \vec{\xi}\cdot\vec{\xi} )=2N_{\S}Q+Q^{2}+{Y^{\parallel}}^{2} \leq 0$. This can only happen at a point where
$Q>0$ (i.e. on $U$) provided $N_S < 0$ there. Moreover, if at any point $q$ on the boundary $\partial U$
we have $N_S |_q=0$, then necessarily the full vector $\vec{\xi}$ vanishes at this point. This implies,
in particular, that the geometric construction of $S_t$ has the property that $q$ remains invariant.

Having noticed these facts, we will now argue by contradiction, i.e. we will assume that
there exists a connected component $U_{0}$ of $U$ 
such that $d(\vec{\xi}\cdot\vec{l}) |_{\partial U_0}\neq 0$ everywhere. In this
circumstances, we can follow the same steps as in the proof of Theorem \ref{thrnonelliptic} to show that,
for small enough $t$ the surface $S_t$ has a portion $S^{+}_t$ lying in the exterior of $S$ and which,
in the Gaussian coordinates above, is a graph over a subset $U_t$ with is a continuous deformation 
of $U_0$. Moreover, the boundary of $U_t$ is a smooth embedded curve $\gamma_t$. The only difficulty
with this construction is that we cannot use $N_S W = Z  <0$ everywhere on $\overline{U}_0$,
in order to conclude that $\theta^{+}_{l, t} < 0$, as we did before.
The reason is that there may be points on $\partial U_0$ where $N_S=0$. However, 
as already noted, these points have the property that {\it do not move at all} by the construction of
$S_{t}$, i.e. the boundary $\gamma_t$ (which is the intersection of $S$ and $S^+_t$)
can only move outside of $U_0$ at points  where $N_S$ is strictly negative. Hence
on the interior
points of $U_t$ we have $N_S <0$ everywhere, for sufficiently small $t$.
Consequently the first order terms in the variation of $\theta_l$, namely
$Z = N_s W$,  is strictly negative on all the interior points of $U_t$. This implies that
$S^{+}_{t}$ has negative outer null expansion everywhere except possibly on its boundary $\gamma_t$.
By continuity, we conclude $\theta^{+}_{l, t} \leq 0$ everywhere. We can now apply Lemma \ref{lemasmoothness}
to $S^{-}_t \cup \gamma_t \cup S^{+}_t$ (where, as before, $S^{-}_t$ is the complementary of $U_t$ in $S$)
to construct a smooth weakly outer trapped surface outside the locally outermost MOTS $S$. This gives
a contradiction. Therefore,  there exists $p\in\partial U_0$ such that $d(\vec{\xi}\cdot\vec{l})|_{p}=0$,
as claimed. $\hfill \square$

{\bf Remark} The assumption $\left. dQ \right|_{ \gamma}\neq 0$ is a technical requirement for
Lemma  \ref{lemasmoothness}.  This is why we had to 
include an assumption on $d Q |_{\gamma}$ in Theorem \ref{thrnonelliptic}  and also that the conclusion
of Theorem \ref{shear2} is stated in terms of the existence of critical points for $Q$. 
If Lemma \ref{lemasmoothness} could be strengthened so as to remove
this requirement, then Theorem \ref{shear2} could be rephrased as stating that any
outermost MOTS in a region where there is a causal Killing vector (irrespective
of its future or past character) must have at least one point where the shear and the 
energy ``density'' along $\vec{l}$ vanish simultaneously. 

In any case, the existence of critical points for a function in the boundary of {\it every}
connected component of $\{ Q >0 \}$ and {\it every} connected 
component of $\{ Q <0 \}$ is obviously a highly non-generic situation. So, locally outermost
MOTS in regions where there is a causal Killing vector or homothety can at most occur under 
very exceptional circumstances.

\section*{Acknowledgments}

We are grateful to M. S\'anchez, J.M.M. Senovilla
and W. Simon for useful comments on the manuscript.
We acknowledge financial support under the projects
FIS2006-05319 of the Spanish MEC, SA010CO5 of the Junta de Castilla y Le\'on 
and P06-FQM-01951 of the Junta de Andaluc\'{\i}a.
AC acknowledgments a Ph.D. grant (AP2005-1195) from the Spanish MEC.

\bibliographystyle{cj}
\bibliography{alberto}{}

\begin{thebibliography}{10}

\bibitem{Dafermos05}
Dafermos, M. (2005) Spherically symmetric spacetimes with a trapped surface.
  {\em Class. Quantum Grav.\/}, {\bf 22}, 2221--2232.

\bibitem{BS09}
Bengtsson, I. and Senovilla, J. The boundary of the region with trapped
  surfaces in spherical symmetry. {\em in preparation\/}.

\bibitem{S08}
Senovilla, J. On the boundary of the region containing trapped surfaces. {\em
  arXiv: 0812.2767\/}.

\bibitem{CollHS01}
Coll, B., Hildebrandt, S., and Senovilla, J. (2001) Kerr-schild symmetries.
  {\em Gen. Rel. Grav.\/}, {\bf 33}, 649--670.

\bibitem{MS03}
Mars, M. and Senovilla, J. (2003) Trapped surfaces and symmetries. {\em Class.
  Quantum Grav.\/}, {\bf 20}, L293--L300.

\bibitem{S03b}
Senovilla, J. (2003) On the existence of horizons in spacetimes with vanishing
  curvature invariants. {\em J. High Energy Physics\/}, {\bf 11}, 046.

\bibitem{AG05}
Ashtekar, A. and Galloway, G. (2005) Some uniqueness results for dynamical
  horizons. {\em Adv. Theor. Math. Phys.\/}, {\bf 9}, 1--30.

\bibitem{AMS08}
Andersson, L., Mars, M., and Simon, W. (2008) Stability of marginally outer
  trapped surfaces and existence of marginally outer trapped tubes. {\em Adv.
  Theor. Math. Phys.\/}, {\bf 12}, 853--888.

\bibitem{Miao05}
Miao, P. (2005) A remark on boundary effects in static vacuum initial data
  sets. {\em Class. Quantum Grav.\/}, {\bf 22}, L53--L59.

\bibitem{CM08}
Carrasco, A. and Mars, M. (2008) On marginally outer trapped surfaces in
  stationary and static spacetimes. {\em Class. Quantum Grav.\/}, {\bf 25},
  055011.

\bibitem{KH97}
Kriele, M. and Hayward, S. (1997) Outer trapped surfaces and their apparent
  horizon. {\em J. Math. Phys.\/}, {\bf 38}, 1593--1604.

\bibitem{S07}
Senovilla, J. (2007) Classification of spacelike surfaces in spacetime. {\em
  Class. Quantum Grav.\/}, {\bf 24}, 3091--3124.

\bibitem{AMS05}
Andersson, L., Mars, M., and Simon, W. (2005) Local existence of dynamical and
  trapping horizons. {\em Phys. Rev. Lett.\/}, {\bf 95}, 111102.

\bibitem{BEM88}
Beem, J., Ehrlich, P., and Markvorsen, S. (1988) Timelike isometries and
  killing fields. {\em Geom. Dedicata\/}, {\bf 26}, 247--258.

\bibitem{Wald84}
Wald, R. (1984) {\em General Relativity\/}. Chicago University Press.

\bibitem{S97}
Senovilla, J. (1997) Singularity theorems and their consequences. {\em Gen.
  Rel. Grav.\/}, {\bf 29}, 701--848.

\bibitem{AMMS09}
Andersson, L., Mars, M., Metzger, J., and Simon, W. (2009) The time evolution
  of marginally trapped surfaces. {\em arXiv: 0811.4721\/}.

\bibitem{AM07}
Andersson, L. and Metzger, J. (2007) The area of horizons and the trapped
  region. {\em arXiv: 0708.4252\/}.

\end{thebibliography}

\end{document}